\documentclass [12pt,elsart,epsfig]{article}     
\usepackage{epsfig}
\begin{document}

\begin {center}
{\large \bf Data on $\bar pp \to \eta ' \pi ^0 \pi ^0 $ for masses
1960 to 2410 MeV/c$^2$}
\vskip 5mm

{A.V. Anisovich$^c$, C.A. Baker$^b$, C.J. Batty$^b$, D.V. Bugg$^a$,  C. Hodd$^a$,
V.A. Nikonov$^c$, A.V. Sarantsev$^c$, V.V. Sarantsev$^c$, 
B.S.~Zou$^{a}$ \footnote{Now at IHEP, Beijing 100039, China} \\
{\normalsize $^a$ \it Queen Mary and Westfield College, London E1\,4NS, UK}\\
{\normalsize $^b$ \it Rutherford Appleton Laboratory, Chilton, Didcot OX11 0QX,UK}\\
{\normalsize $^c$ \it PNPI, Gatchina, St. Petersburg district, 188350, Russia}\\ 
[3mm]}
\end {center}

\begin{abstract}
Data on $\bar pp \to \eta '(958)\pi ^0\pi ^0$ are presented at nine $\bar p$
momenta from 600 to 1940 MeV/c.
Strong S-wave production of $f_2(1270)\eta '$ is observed, requiring a
$J^{PC} = 2^{-+}$ resonance with mass $M = 2248 \pm 20$ MeV,
$\Gamma = 280 \pm 20$ MeV.
\end{abstract}

The first data are presented on $\bar pp \to \eta '\pi ^0 \pi ^0 $
in flight.
These data were taken with the Crystal Barrel detector at LEAR.
They are part of an extensive study of the $I = 0$, $C = +1$
system in several channels.
Data have been reported earlier on $\pi ^0\pi ^0$ [1], $\eta \eta$ and
$\eta \eta '$ [2], and $\eta \pi ^0 \pi ^0$ [3].
A comparison will be made here specifically with the
$\eta \pi ^0 \pi ^0$ data,
and with a combined amplitude analysis of all the earlier data [4].

The experimental set-up has been reported in detail [5].
A $\bar p$ beam from LEAR interacts in a liquid hydrogen target 4.4 cm
long at the centre of the detector.
Incident $\bar p$ are counted by a coincidence between a scintillator
of 5 mm diameter and a small multiwire proportional chamber, both positioned
$\sim 5$ cm upstream of the target.
Two veto counters 20 cm downstream of the target provide a trigger
for interactions.
The target is surrounded over 98\% of the solid angle by a multiwire
proportional chamber and a silicon vertex detector.
These provide an on-line trigger for neutral final states.
With a $\bar p$ beam of $\sim 2 \times 10^5$/s, the trigger rate is
$\sim 60$/s.

The present channel is studied in $10\gamma$ events, where $\eta ' \to
\eta \pi ^0\pi ^0$, $\eta \to \gamma \gamma$.
Photons are detected with high efficiency down to 20 MeV in a barrel
of 1380 CsI crystals covering 98\% of the solid angle; 
the geometry is such that crystals point towards the target.
The crystals have a length of 16 radiation lengths  and provide an angular
resolution of $\pm 20$ mrad in azimuth and polar angle.
The energy resolution is given by $\Delta E/E = 2.5\%/E$(GeV)$^{1/4}$.

The general procedures for event reconstruction and selection have
been described in several earlier publications, of which the most detailed
concern the study of $\pi ^0 \pi ^0$, $\eta \eta$ and $\eta \eta '$
final states [1,2].
A Monte Carlo simulation of the detector is used to assess the
efficiency for reconstruction of the $\pi ^0 \pi ^0 \eta '$ final state
and the levels of background from competing channels.

\begin{figure}
\begin{center}
\vskip -10mm
\epsfig{file=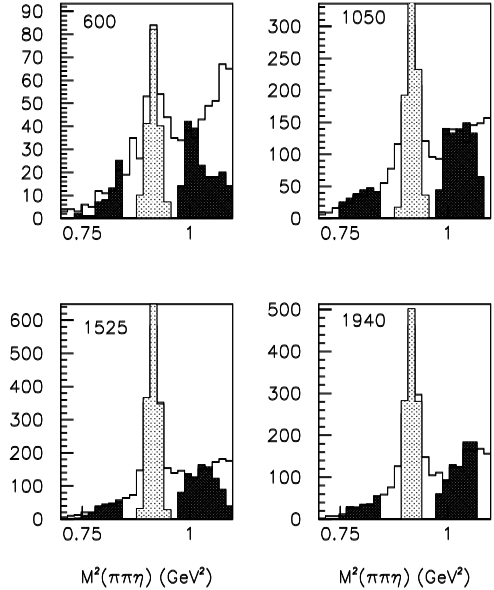,width=8cm}\
\vskip -91.95mm
\epsfig{file=fig1B_eppp.ps,width=8cm}\
\caption{The distribution of $M^2(\eta \pi \pi)$ at four beam momenta
indicated by numerical values in MeV/c. The shaded areas show selected
signal events in the $\eta '$ peak and those used for sideband subtraction. }
\end{center}
\end{figure}

Events are first submitted to a kinematic fit to $\bar pp \to 10\gamma$,
requiring a confidence level $>5\%$.
The  best kinematic fit to $\bar pp \to \eta 4\pi ^0$ is then selected, again
with confidence level $>5\%$.
At this step, the main background to $\eta 4\pi ^0$ 
comes from $5\pi ^0$ events.
This background is suppressed strongly by rejecting any event passing
a kinematic fit to
$5\pi ^0$ with confidence level $>1\%$ (or 0.1\% at
600 MeV/c, where the background is more severe).
Finally, those few events are rejected which fit $\eta \eta 3\pi ^0$ with
confidence level better than $\eta 4\pi ^0$.

Fig. 1 illustrates at four beam momenta the $\eta \pi \pi$ mass 
distribution of surviving events in the mass range around the $\eta '$.
There is a clear $\eta '$ signal, agreeing in mass within
$\le 4$ MeV with the standard value at all momenta.
It is superposed on a smooth background, whose magnitude is
largest at low beam momenta.
The Monte Carlo simulation estimates that the background comes
approximately equally from 3 sources: (i) $5\pi ^0$ events, (ii)
$\omega 4\pi ^0$, $(\omega \to \pi ^0 \gamma )$ after losing one photon,
and (iii) $\eta 4\pi ^0$ without an $\eta '$.
The predicted background agrees with that observed (within 10\% of the
prediction).
Tighter cuts do not improve the signal/background ratio significantly,
but simply cause loss of events.

Signal events are selected from the peak region of the $\eta '$ by
adjusting a mass cut around the peak at every individual momentum
so as to optimise the signal/background ratio.
Very rarely, two events fall within the window; in this case the one
closer to the $\eta '$ is accepted.
Statistics of the data selection are shown in Table 1.
In the maximum likelihood fit used for amplitude analysis, sidebins 
events shown shaded in Fig. 1 are used to subtract the background.
The areas of sidebins are chosen so that each covers twice the range of
mass squared which is used to select $\eta '$ events; in this way, 
statistical errors on the background are small.
A technicality is that the width of the mass cut is varied according to 
the accuracy with which
the $\eta '$ mass is reconstructed. This is the reason that
sidebands have diffuse edges: the width of the sidebin likewise varies with the
width of the $\eta '$ mass cut.
Technically, the way the subtraction is made is to include sidebin events
into the fit with a weight $-0.25$ times that of events selected in
the signal region.
Amplitudes are constructed with tensor expressions using the
measured mass of each $\eta '$.

\begin{figure}
\begin{center}
\vskip -35mm
\epsfig{file=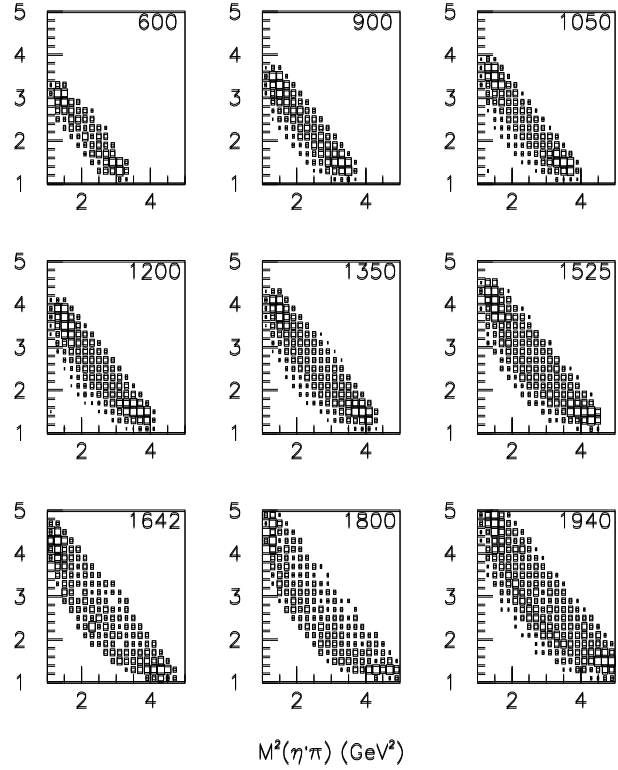,width=11cm}\
\vskip -123.72mm
\epsfig{file=dalitz_eppp.ps,width=11cm}\
\vskip -4mm
\caption{Dalitz plots at all beam momenta for events from the signal
region of Fig. 1. Numerical values indicate beam momenta in Mev/c. }
\end{center}
\end{figure}
\begin{figure}
\begin{center}
\vskip -35mm
\epsfig{file=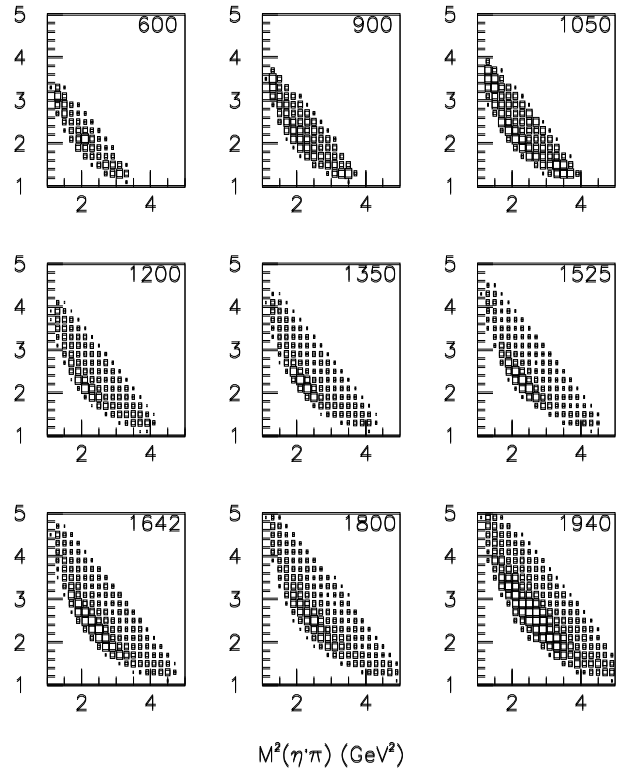,width=11cm}\
\vskip -123.72mm
\epsfig{file=dalitz_back.ps,width=11cm}\
\vskip -4mm
\caption{Dalitz plots for events from the sidebin regions of Fig. 1.}
\end{center}
\end{figure}

Fig. 2 shows the Dalitz plots at all momenta for events from the
signal region.
There is an obvious contribution due to $f_2(1270)\eta '$,
appearing at momenta $\ge 1200$ MeV/c at the lower left edge of the plot.
Fig. 3 shows the Dalitz plots for sidebin events.
The distribution of background is not uniform, but peaks in the
corners of the Dalitz plots.
This peaking accounts for corresponding peaks observed in the
corners of the Dalitz plots of Fig. 2.
When the subtraction is made, the surviving signal outside the
$f_2(1270)$ peak is nearly uniform within the available
statistics.
At 1940 MeV/c, there is also some weak $f_2(1270)$ in the background; 
we have checked that this is not due to $\eta '\pi ^0 \pi ^0$ signal
spilling into the mass ranges used for the sidebins.

\begin{table} [htbp]
\begin{center}
\begin{tabular}{ccccc}
\hline
Momentum & Data & BG & Signal & $\epsilon$ \\
(MeV/c)  &      &    &        & (\%) \\\hline
600  &  180 &  61 &  119 & 2.90\\
900  & 1017 & 399 &  618 & 4.61  \\
1050 &  831 & 257 &  574 & 5.76 \\
1200 & 2770 & 852 & 1918 & 6.33 \\
1350 & 2296 & 595 & 1701 & 5.92\\
1525 & 1416 & 381 & 1035 & 5.06\\
1642 & 1530 & 330 & 1200 & 4.72\\
1800 & 1503 & 325 & 1178 & 4.57\\
1940 & 1063 & 240 &  823 & 4.34\\\hline
\end {tabular}
\caption {Numbers of selected events, estimated background (BG),
true signal, and reconstrucion efficiency $\epsilon$ as a function of
beam momentum.}
\end{center}
\end{table}
There is no indication for the presence of $a_2(1320) \to \eta '\pi$.
The expected contribution may be predicted from fits which have been
made to $a_2(1320)\pi$ in $\eta \pi ^0 \pi ^0$ data [3].
The predicted contribution is only $\sim 3\%$ of $\eta '\pi ^0 \pi ^0$,
because of the small (0.53\%) branching fraction of $a_2(1320)$ to
$\eta '\pi$.
This contribution is included in the amplitude analysis
using amplitudes fitted to the $\eta \pi \pi$ data,
but is so small as to have negligible effect on conclusions.
Fig. 4 shows projections at two beam momenta on to masses of $\pi\pi$
and $\pi \eta$; the latter is featureless.
The histograms show results of the maximum likelihood fit described below.
\begin{figure} [h]
\begin{center}
\vskip -12mm
\epsfig{file=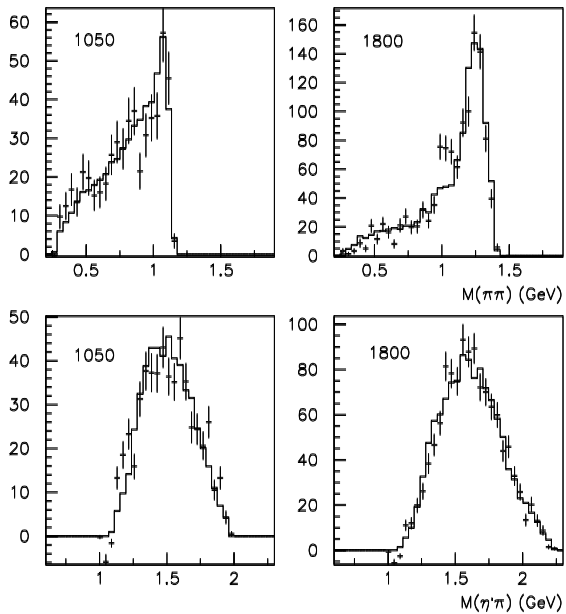,width=9cm}\
\vskip -95.735mm
\epsfig{file=HIST1.PS,width=9cm}\
\vskip -8mm
\caption {Projections on to $M (\pi \pi )$ and $M (\eta '\pi)$ at 
beam momenta of 1050 and 1800 MeV/c; in all cases, a background subtraction is
made using sidebins. 
Histograms show the fit compared with data.}
\end{center}
\end{figure}

We now turn to physics results. 
Data points on Fig. 5(a) show the integrated $\eta '\pi ^0\pi ^0$
cross section after background subtraction and after scaling to allow 
for all other unobserved decay modes of $\eta '$, $\eta$ and $\pi ^0$.
There is a peak around 2230 MeV, which is the nominal threshold for
$f_2(1270)\eta '(958)$.
The absolute normalisation is obtained using beam counts, target
length and density,
and correcting the observed number of signal events for the
reconstruction efficiency shown in Table 1.
A correction is applied for observed dependence of the
cross section on beam rate, as described in detail in Ref. [1].
\begin{figure}
\begin{center}
\epsfig{file=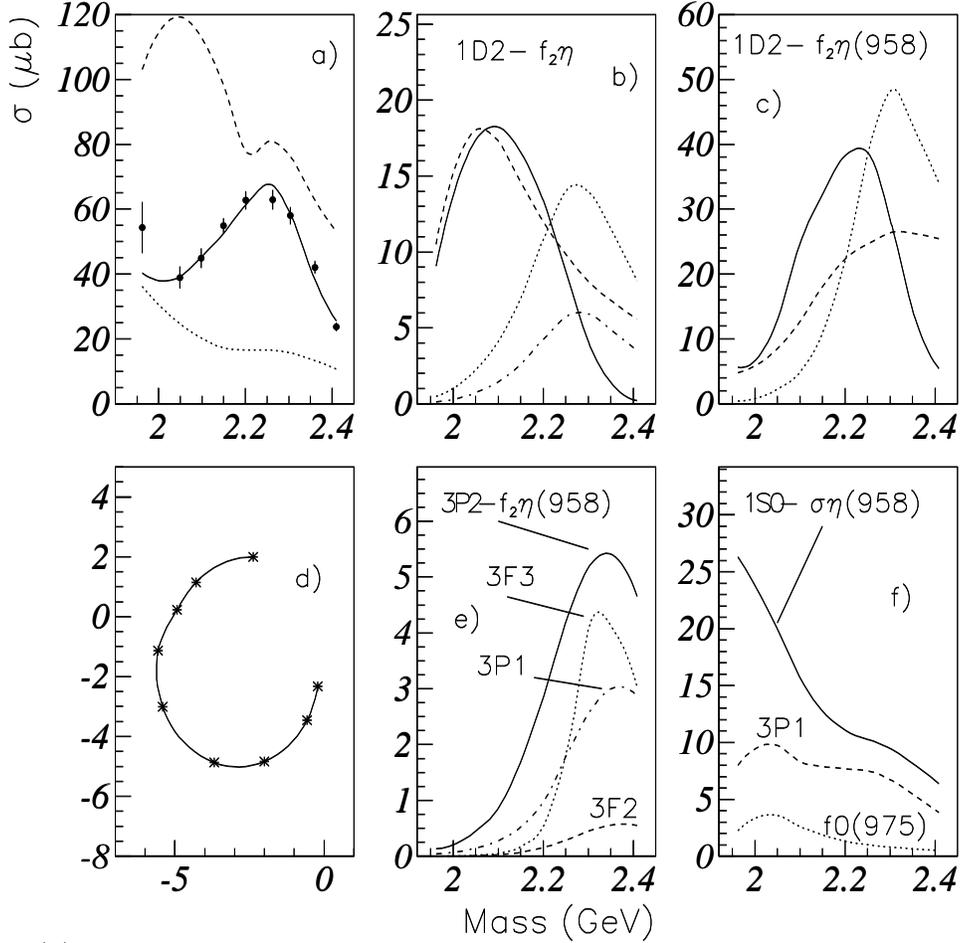,width=14cm}\
\vskip -14.025cm
\epsfig{file=HIST3.PS,width=14cm}\
\vskip -8mm
\caption{(a) Points with errors show the integrated cross sections for
the final state $\eta '\pi ^0\pi ^0$, after correction for backgrounds and
for all decay modes of $\eta '$, $\eta$ and $\pi ^0$;
the full curve shows the fit from the amplitude analysis;
the dashed curve shows the $\eta \pi \pi$ cross section from Ref. [3],
multiplied by the SU(3) factor $(0.75)^2$;
the dotted curve shows the $\sigma \eta '$ contribution;
(b) the full curve shows the cross section for $f_2\eta$ fitted to
$\eta \pi \pi$ data; the dotted curve shows the contribution to $\eta \pi \pi$
from $\eta _2(2248)$ alone and the dashed curve that
from $\eta _2(1860) + \eta _2(2030)$; the chain curve shows the intensity
fitted to $f_2(1270)\eta$ with $L = 2$ in Ref. [4];
(c) as (b) for $f_2\eta '$;
(d) Argand diagram for the $f_2\eta '$ S-wave amplitude; crosses mark
beam momenta; (e) Intensities of contributions to $f_2\eta '$ from $^3P_2$
(full curve), $^3P_1$ (chain curve), $^3F_3$ (dotted) and $^3F_2$ (dashed);
(f) intensities of contributions to $\sigma \eta '$ from $0^-$ (full
curve), $1^+$ (dashed) and $1^+ \to f_0(975)\eta '$ (dotted).}
\end{center}
\end{figure}

The amplitude analysis is made using (a) S and P-waves for
$\sigma \eta '$, where $\sigma$ stands for the $\pi \pi$ S-wave
amplitude, for which we use the parametrisation of Zou and Bugg [6],
(b) S and P-waves for $f_2(1270)\eta '$, and
(c) a small, almost negligible contribution from $^3P_1 \to f_0(975)\eta '$,
which helps marginally in fitting the $\pi \pi$ mass distribution at the
lowest three beam momenta.
It is to be expected that higher partial waves for $f_2\eta '$ will be
suppressed strongly by the centrifugal barrier in the final states.
Contributions from $f_2\eta '$ D-waves have been tried in the fit, 
but are not required;
indeed, the P-wave contribution is quite small.
Likewise, $\sigma \eta '$ contributions with $L \ge 2$ are negligible.

We shall present amplitudes for $f_2(1270)\eta '$ in partial waves
$^1D_2$ $(J^{PC} = 2^{-+})$, $^3P_2$ and $^3F_2$ $(2^{++})$,
$^3P_1$ $(1^{++})$ and $^3F_3$ $(3^{++})$; they will be compared with
$f_2(1270)\eta$ observed in $\eta \pi\pi$ data [3,4].
These two channels are related by the composition of the $\eta '$ and
$\eta$ in terms of strange and non-strange quarks:
\begin {eqnarray}
|\eta >  &\simeq& 0.8 \frac {u\bar u + d\bar d}{\sqrt {2}} - 0.6 s\bar s,\\
|\eta '> &\simeq& 0.6 \frac {u\bar u + d\bar d}{\sqrt {2}} + 0.8 s\bar s.
\end {eqnarray}
The coefficients 0.8 and 0.6 are derived from the well known  pseudo-scalar
mixing angle [7].
Our earlier analysis of
$\bar pp \to \pi ^-\pi ^+,$ $\pi ^0 \pi ^0$,
$\eta \eta$ and $\eta \eta '$ [8] finds that almost all $s$-channel
resonances produced in $\bar pp$ interactions are consistent with small
mixing angles $\le 15^{\circ}$ between $(u\bar u + d\bar d)/\sqrt {2}$ and
$s\bar s$.
The naive prediction is therefore that amplitudes $a$ for $\bar pp \to
f_2(1270)\eta '$ and $\bar pp\to f_2(1270)\eta $ will be related by
\begin {equation}
a(f_2\eta ') \simeq 0.75 a(f_2\eta ).
\end {equation}

The peak in the full curve of Fig. 5(a) requires a resonance in 
$f_2\eta '$ close to the mass of the peak.
However, the mass spectrum from a simple resonance will be pushed
upwards by the rapidly increasing phase space for the final state
$f_2\eta '$.
This effect is visible in the dotted curve of
Fig. 5(c), which shows the resonance contribution to $f_2\eta '$ fitted to
$\eta _2(2248)$; this curve peaks above 2300 MeV because of the increasing
phase space.
In order to reproduce the integrated cross section of
Fig. 5(a), the amplitude analysis requires a strong interfering background
peaking below threshold.
The interference
is constructive at  low masses, and is required to give a
large $f_2\eta '$ cross section there, despite the limited phase space.
Above the peak at 2230 MeV, the interference becomes destructive,
and cuts off the $f_2\eta '$ cross section on the upper side of the
resonance.

The motivation for including this background contribution at low $f_2\eta '$
masses arises from the new combined analysis [4] of $\eta \pi \pi$ data, 
together with those on $\bar pp \to \pi ^-\pi ^+,$ $\pi ^0 \pi ^0$,
$\eta \eta$ and $\eta \eta '$.
Results for $\eta \pi \pi$ from that analysis are shown in Fig. 5(b).
That analysis requires a $2^{-+}$ resonance at
$2267 \pm 14$ MeV. It appears there most clearly in $f_2(1270)\eta $
with $L = 2$ in the final state, shown by the
chain curve in Fig.5(b).
However, for the dominant $f_2 \eta$ $L = 0$  channel, what one observes is
a strong peak near 2 GeV, shown by the full curve.
This comes mostly from $\eta _2(1860)$,
but partly from $\eta _2(2030)$ reported in an analysis of data on
$\bar pp \to \eta \pi ^0 \pi ^0 \pi ^0$ [9].
The intensities of contributions to the $f_2\eta$ channel are shown
in Fig. 5(b)  from (i) all $\eta _2(2248)$ contributions (dotted curve) and
(ii) the coherent sum of $\eta _2(1860)$ and $\eta _2(2030)$ (dashed curve);
the latter two
resonances are not well resolved by the $\eta \pi \pi$ data, because they
lie close together near the $\bar pp$ threshold.
The contribution from $\eta _2(2248)$ interferes destructively with $\eta
_2(1860)$ and $\eta _2(2030)$, so as to cut off the full curve at high masses.

In present data, the width of the $\eta _2(2248)$ is well determined by 
the width of the peak in Fig. 5(a): $\Gamma = 280 \pm 20$ MeV.
This determination is superior to that in $\eta \pi \pi$ data: $290 \pm 50$
MeV.
The mass is somewhat less well determined, since the interference
with the tails of the lower resonances may shift the peak by an amount
which is sensitive to  their widths.
Using the best estimates for the widths from Ref. [8], the mass 
from the present data is $M = 2248 \pm 20$ MeV,
in reasonable agreement with the value derived from $\eta \pi \pi$ data:
$2267 \pm 14$ MeV.
The Argand diagram for the $f_2\eta '$ S-wave amplitude is
shown in Fig. 5(d).

A striking feature of the $f_2\eta '$ signal is its large magnitude.
The dashed curve on Fig. 5(a) shows the {\it complete} 
integrated $\eta \pi ^0\pi ^0$
cross section, multiplied by $(0.75)^2$ to allow for the expected
inhibition  of $\eta '$ with respect to $\eta$.
It is surprising that the $f_2\eta '$ signal is nearly as strong as the
dashed curve, bearing in mind the difference in available phase space for
$f_2\eta '$ and $f_2\eta$.
The peak in the $\eta ' \pi \pi$ cross section (full curve) is much
larger than the small peak observed at the same mass in the
$\eta \pi \pi$ cross section.
Likewise, the S-wave peak due to $\eta _2(2248) \to f_2\eta '$,
shown by the dotted curve in Fig. 5(c), is
considerably stronger than that in $f_2\eta$ in Fig. 5(b).
If one takes into account  the available phase space for
$f_2\eta '$ and $f_2\eta$, the coupling constant for
$\eta _2(2248) \to f_2\eta '$ relative to that in $f_2\eta$ is stronger
than predicted by equn. (3) by a factor $5.2$ in amplitude.

Vandermeulen has remarked that $\bar pp$ annihilation usually
favours high mass final states [10]. This may be understood as a
form factor effect, arising from the sizes of the participating states.
In present data, the final state $f_2\eta '$ has very low momentum.
However, in the process $\eta _2(2248) \to f_2\eta$, the momentum $q$
in the final state is $\sim 635$ MeV/c.
The factor 5.2 would require a form factor
$\exp -(4.1q^2)$ in amplitude, with $q$ in GeV/c;
if this arises from a source
having a Gaussian distribution in $r$, the form factor takes the
well known form $\exp -(q^2R^2/6)$, and requires a
radius of interaction $R = 0.98$ fm.
Such a form factor is surprisingly strong. For comparison,
the Vandermeulen form factor approximates to $\exp -(1.5q^2)$.

A possibility is that $\eta _2(2248)$ is an $s\bar s$ state.
However, strong production from $\bar pp$ is unlikely and in disagreement
with results for $\pi \pi $, $\eta \eta $ and $\eta \eta '$ [4].

The strong sub-threshold contribution to the $f_2\eta '$ S-wave is
intriguing.
A variety of explanations are possible, of which we mention one.
In Ref. [9], evidence has been presented for three $\eta _2$
resonances in a mass range where only two are likely to be
$q\bar q$. Of these, $\eta _2(1860)$ is a candidate for a hybrid,
because of its strong decay to $f_2\eta$, despite limited phase space.
If that conjecture is correct, it should be accompanied by an
$s\bar sg$ partner at about 2100 MeV. Such an $s\bar sg$ hybrid
is expected to decay strongly to $f_2(1525)\eta '$ and $f_2(1270)\eta '$.
If it mixes into neighbouring $q\bar q$ states, it could help to
explain the anomalously strong $f_2\eta '$ signal observed here.

\begin{figure}
\begin{center}
\vskip -25mm
\epsfig{file=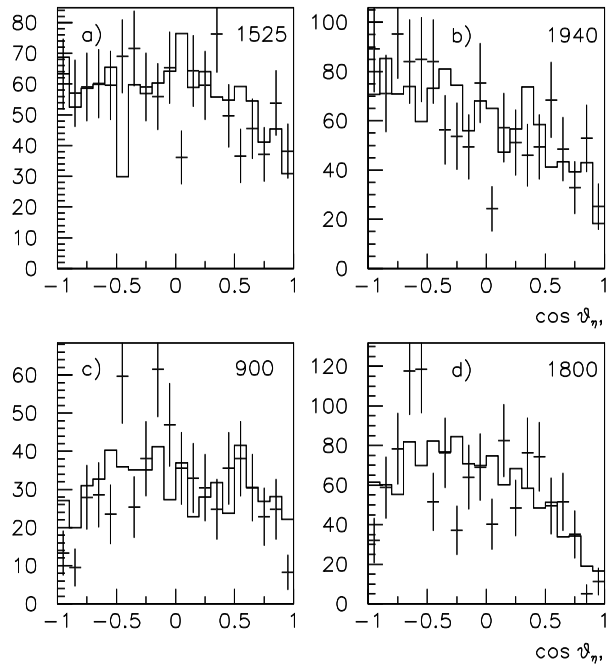,width=9.9cm}\
\vskip -106mm
\epsfig{file=HIST2.PS,width=9.9cm}\
\caption{Production angular distributions for $f_2\eta '$ at
(a) 1525 and (b) 1940 MeV/c for $M(\pi \pi ) >1.1$ GeV;
also for $\sigma \eta '$ at
(c) 900 and (d) 1800 MeV/c for $M(\pi \pi ) <1$ GeV.
Points with errors show data, uncorrected for
acceptance; histograms show the maximum likelihood fit.}
\end{center}
\end{figure}

We now consider other partial waves.
The present data require a small but significant P-wave $f_2\eta '$
contribution.
This could arise from initial $\bar pp$ states $^3P_1$, $^3P_2$,
$^3F_2$ or $^3F_3$.
The amplitude analysis of Ref. [4] requires all of these contributions
in $\eta \pi \pi$ data with a $3^+$ resonance at 2303 MeV, a $1^+$ 
resonance at $2310 \pm 60$ MeV and $2^+$ resonances at 2240 and 2293 MeV.
A good fit to present data may be obtained by fixing the relative
magnitudes and phases of these partial waves from the fit to
$\eta \pi ^0 \pi ^0$ data.
The absolute magnitude of the P-wave contribution is  sensitive
to the radius chosen for the Blatt-Weisskopf centrifugal barrier. 
This radius is
therefore adjusted to give the best fit to the data, with the
reasonable result $0.8$ fm.

The magnitudes of the contributions are
then 3.5\% for $^3F_3$, $3.2\%$ for $^3P_1$, $3.2\%$ for the $2^+$
resonance at 2240 MeV and 1.0\% for the $2^+$ resonance at
2293 MeV; in the latter two, the ratios of amplitudes for
$^3P_2$ and $^3F_2$ are taken from Ref. [4].
Without these amplitudes, log likelihood of the fit to $\eta '\pi ^0 \pi ^0$
is worse by 142 for only one parameter fitted to the overall magnitude;
so the P-wave contribution is highly significant.
[Our definition of log likelihood is such that it a change of 0.5
corresponds to one standard deviation change in one variable].
If instead the magnitudes and phases of these amplitudes are fitted
freely,
the fit changes very little. It is not possible from the present data
to separate $^3P_2$ and $^3F_2$, which need to be constrained in
relative magnitude as determined in Ref. [4].
With this constraint, the freely fitted intensities are $3.9\%$ for $^3F_3$, 
$4.7\%$ for $^3P_1$ and $3.9\%$ for $2^+$, close to the contrained fit.

Figs. 6(a) and (b) show angular distributions 
for production of $\eta 'f_2(1270)$ in the mass range $>1.1$ GeV 
in terms of the centre of mass angle $\theta$ of the $\eta '$ 
The distributions are
uncorrected for acceptance, which is included in the maximum likelihood
fit shown by the histograms.
At high beam momenta, the acceptance for $\eta '$ falls in the
forward direction, where
the separation of its decay products becomes less efficient.
A check on the reconstruction procedure is that angular distributions
are symmetric forward-backward in the centre of mass system within errors,
after correction for acceptance; this symmetry is 
required by charge conjugation invariance.

We now turn to the contributions from the broad $\sigma \eta '$
channel.
From present data, the only firm conclusion which may be drawn is that
contributions from both $^1S_0$ and $^3P_1$ initial states are required.
At all momenta from 900 MeV/c upwards, the data require angular
distributions of the form $A + B\cos ^2\theta _{\eta '}$, as shown
in Figs. 6(c) and (d).
The $\eta \pi \pi$ data have been fitted including $0^-$ and $1^+$
resonances. Present data are fitted well by the same resonances.
However, statistics are not sufficient to provide clear evidence
of these resonances in present data.
Fig. 6 shows that the fit to  data is adequate.

In summary, the main feature of the $\eta '\pi ^0\pi ^0$ data
is a peak at 2230 MeV, requring a dominant contribution from
the $f_2(1270)\eta '$ S-wave. The data require a $2^{-+}$
resonance with mass $2248 \pm 20$ MeV and width $\Gamma = 280 \pm 20$
MeV; this result is closely consistent with an $\eta _2(2267)$
resonance observed in $\eta \pi \pi$ data. The $f_2\eta '$ S-wave
amplitude is surprisingly strong compared with that for $f_2\eta$,
even allowing for a form factor in the latter.
Contributions from $f_2(1270)\eta '$ P-states are consistent
with the amplitude analysis of the $\eta \pi \pi$ data.

\section{Acknowledgement}
We thank the Crystal Barrel Collaboration for allowing use of the data.
We wish to thank the technical staff of the LEAR machine group and of all
participating institutes for their invaluable contributions to the successful
running of the experiment. We acknowledge financial support from the
British Particle Physics and Astronomy Research Council (PPARC).
The St. Petersburg group wishes to acknowledge financial support from PPARC and
INTAS grant RFBR 95-0267.

\end {document}